\documentclass[aps,pre,showpacs,groupedaddress,floatfix]{revtex4}
\usepackage{graphicx}
\begin{document}

\title{Dynamical heterogeneity and jamming in glass-forming liquids}   
    
\author{Naida La\v{c}evi\'{c}$^{a,}$\footnote{New address: Department of Chemical Engineering, University of California Berkeley, Berkeley, CA 94706, USA} and Sharon C. Glotzer$^{a,b}$ }

\affiliation{${}^a$Department of Chemical Engineering and ${}^b$Department of Materials Science and Engineering \\ University of Michigan, Ann
Arbor, MI 48109, USA}

\date{May $27$, $2004$  }

\begin{abstract} 

The relationship between spatially heterogeneous dynamics (SHD) and
jamming is studied in a glass-forming binary Lennard-Jones system via
molecular dynamics simulations.  It has been suggested by O'Hern et
al.~\cite{O'Hern200186111} that the probability
distribution of interparticle forces $P(F)$ develops a peak at the
glass transition temperature $T_g$, and that the large force
inhomogeneities, responsible for structural arrest in granular
materials, are related to dynamical heterogeneities in supercooled
liquids that form glasses. It has been further suggested that ``force
chains'' present in granular materials may exist in supercooled
liquids, and may provide an order parameter for the glass transition.
Our goal is to investigate the extent to which the forces experienced
by particles in a glass-forming liquid are related to SHD, and compare
these forces to those observed in granular materials and other
glass-forming systems. Our results are summarized as follows. We find
no peak in $P(F)$ at any temperature in our system, even below $T_g$.
We also find that particles that have been localized for a long time
are less likely to experience high relative force and that mobile
particles experience higher relative forces at shorter time scales,
indicating a correlation between pairwise forces and particle
mobility. We construct force chains based on the magnitude of pairwise
forces.  We find that force chains constructed in this manner are
composed of both localized and mobile particles, therefore there is no
one-to-one correspondence between force chains as defined here and
locally mobile or immobile regions of the liquid. We also find that
force chains do not play the same role as force chains in granular
materials, but may indicate a difference in the evolution of the local
environment of particles with different mobility. We also discuss a
possible relationship between force chains found here and the
development of string-like motion found in other glass-forming
liquids~\cite{DonatiStringPRL,YeshiJCP2004}.

\end{abstract}


\maketitle

\section{Introduction}
\label{Sec1}
Attempts are underway in the statistical mechanics community to unify
concepts regarding supercooled liquids and granular materials, two
very different classes of systems that display similar behavior in
many respects.  Systems near their glass transition,
colloidal suspensions at large pressures or densities, foams under
shear, granular materials under ``tapping'' or shearing, and other
systems like bubbles and droplets ``jam'' under certain circumstances.
Jamming is also seen in polymer crazes~\cite{Rottler200289}.  Jamming
is a process in which systems appear ``stuck'' in phase space because
their particles come in close contact with each other resulting in
structural arrest. This process sketches what happens
with supercooled liquids at the glass transition temperature $T_g$,
where jamming is controlled by the temperature or density. In
colloidal suspensions under high pressure or at high density,
particles also get ``stuck'', and the suspension acts as an amorphous
solid\cite{Trappe2001411772}. Foams or emulsions flow under high shear
rate, but at low shear they stop flowing and appear to be solids as
well\cite{Durian1995754780}.  Extensive reviews of granular materials
can be found in
Refs.~\cite{deGennes199971S374,Kadanoff199971435,Jaeger1996681259}.
Among many experiments that deal with the behavior of granular
materials under external perturbation,
Refs.~\cite{Nowak1998571971,Knight1995513957,D'Anna20018725} indicate
that the response of granular materials upon isolated tapping or
continuous vibration is similar to the response of supercooled liquids
near $T_g$.  In these experiments, the density fluctuations of a
granular material subject to shaking were investigated by reducing the
shaking frequency until the granular material reached a ``jammed''
configuration.

What are the similarities and differences between glass-forming
liquids and granular materials? First we note some of the properties
of granular materials.  Unlike glass-forming liquids, granular
materials consist of a large number of particles that are individually
solid (grains). The grain-grain interactions are classical because the
size of the grains is much larger than the de Broglie wavelength. The
grains exert forces only when they are in contact, and may be
surrounded by a fluid (typically air) or a vacuum. The collisions
between grains are in general inelastic.  The main difference between
liquids and granular materials, aside from the obvious difference of
length and time scale, is the concomitant difference in energy scales. In
granular materials, thermal energies $k_b T$ are insignificant
compared to the energy it takes to move a single particle. Thermal
energy in the liquid allows it to explore different states, while the
low relative magnitude of thermal energy in a granular material does
not allow it to sample other configurations unless energy is added to 
the system. This means that granular materials can stay in a
metastable state indefinitely.

A schematic ``jamming'' phase diagram was proposed in
Ref.~\cite{Liu199839621} in order to unite the concepts of jamming in
many different systems.  According to the diagram, jamming in
supercooled liquids occurs at low temperature $T$ and high pressure
$P$. Such unified picture suggests several questions, such as whether
the dynamics of different systems approaching the jammed state are
similar. The jamming phase diagram also suggests that the glass and
jammed states may be related. However, the details of this
relationship are not yet fully understood. For instance, a commonly
measured quantity in granular materials is the probability
distribution of normal forces $P(F)$. $P(F)$ develops a peak near the
jamming transition in granular materials that is well established in
experiments and computer simulations.  It has been suggested that a
similar peak in $P(F)$ measured in supercooled liquids signifies the
onset of solidity, and consequently is a signature of the glass
transition.  The obvious supposition is: if the link between jamming
and the glass transition exists, then the mechanism for slowing down
of dynamics in supercooled liquids may be related to macroscopic
structural arrest in granular materials. A peak in $P(F)$ at and below
$T_g$ would provide a link in this regard. Such a peak was reported in
$2$-d simulations of several glass-forming
liquids~\cite{O'Hern200186111}.  We test the generality of this result
in the present work.
 
As shown in many experiments and computer simulations of granular
materials~\cite{Longhi200289,Howell1999825241,Roux2000616802,Radjai199677274,Tkachenko199960687},
large force inhomogeneities are responsible for structural arrest
under shear. If the shear rate is sufficiently low, the granular
material will be structurally ``arrested'' or jammed.  Jamming in
granular materials occurs because the particles form ``force chains''
along the direction in which the stress is
applied~\cite{Farr1997557203}. This leads us to pose the following 
question: Do
force chains exist in supercooled liquids, and if so, are they 
responsible for the slowing down of
dynamics, and related to other prominent features of supercooled
liquids?

A salient feature of glass-forming liquids whose origin is still not
understood is the spatially heterogeneous nature of their dynamics. In
particular, the non-exponential character of the relaxation of density
correlation functions and decoupling of the transport coefficients can
be rationalized with the existence of {\it spatially heterogeneous
dynamics} (SHD) or ``dynamical heterogeneity''. We refer to a system
as dynamically heterogeneous if it is possible to select a dynamically
distinguishable subset of particles by experiment or computer
simulation~\cite{Bohmer19982351}.
Simulations~\cite{heuer-sims,onuki,harrowell,mountain,hiwatari,dzugutov,glotzer-kdppg}
and
experiments~\cite{schmidt-rohr,sillescu,bohmer,ediger,torkelson,heuer-exp,leporini,richert,israeloff,chamberlin,russina,weeks,kegel,wiltzius,vandenbout,marcus}
have demonstrated the cooperative and spatially heterogeneous nature
of the liquid dynamics (for reviews of the experimental evidence for
spatially heterogeneous dynamics, see, e. g.,
Refs.~\cite{sillreview,ediger-rev,richrev}).

In
Refs.~\cite{glotzer-kdppg,bdbg,dgpkp,dgp,DonatiStringPRL,dg,scg-review,LacevicJCP2003},
several approaches -- including calculation of a
displacement-displacement correlation function, identification of
clusters of mobile particles, and calculation of a four-point
time-dependent density correlation function -- predicted and
demonstrated the importance of spatially heterogeneous dynamics in
supercooled liquids using a rigorous statistical mechanical analysis.
In particular, using the four-point time-dependent density correlation
function formalism, we found dynamical correlation lengths of regions
of localized and delocalized particles which suggests a picture of
fluctuating domains of temporarily localized and delocalized
particles, as suggested by Stillinger and
Hodgedon~\cite{stillinger}. It has recently been
demonstrated in two simulated liquids~\cite{magnus} that these fluctuating domains are dynamically facilitated, as predicted by Garrahan and
Chandler~\cite{chandler}.  Many specific predictions made possible
through such analyses have now been confirmed in experiments on
colloidal suspensions~\cite{weeks,kegel,glotzer-science-phworld}. The
tools from these analyses are thus available to investigate the
connection between jamming in granular materials and SHD in
supercooled liquids. In this work, we will use the four-point
time-dependent density correlation function formalism to test this
connection.

Here we analyze the forces experienced by particles in a glass-forming
liquid in analogy to those in granular materials, investigate $P(F)$, 
and attempt to
ascertain a relationship between the forces and spatially heterogeneous
dynamics.  The paper is organized as follows. In Section~\ref{method},
we describe the model and method used to produce our results. In
Section~\ref{summary_SHD_4pnt}, we give a brief description of SHD and
the theoretical framework used to describe SHD. In
Section~\ref{AvgForInst}, we measure the instantaneous forces between
all particle pairs in our glass-forming liquid, and calculate the
average force and corresponding standard deviation for every $T$
simulated. In Section~\ref{sub_force_inst}, we divide the set of
instantaneous pair forces for each configuration into subsets of
highest, average and lowest forces at each $T$.  Using our definition
of localized and replaced particles (defined in
Section~\ref{summary_SHD_4pnt}), we find the fractions of these
particles in each subset of instantaneous pair forces. In
Section~\ref{force_chains}, we define ``paths'' of highest, average
and lowest forces, which we call ``force chains'', and investigate the
temperature dependence of their average mass, average number, and
distribution. In Section~\ref{frac_loc_deloc}, we investigate the
fractions of localized and replaced particles in these ``force
chains''.  We conclude with the discussion of our results in
Section~\ref{discussions}.

\section{Model and method}
\label{method}

The simulation method we use to generate data for our analyses is
molecular dynamics ($\rm{MD}$). This is a widely used method in the
investigation of supercooled liquids and glasses that provides static
and dynamic properties for a collection of particles described by
classical force fields. To perform our simulations we use
LAMMPS~\cite{lammps}, a parallel MD code developed by Plimpton.
We study a $50/50$ binary mixture of particle types ``$A$'' and
``$B$'' that interact via the Lennard-Jones potential
\begin{eqnarray} 
V_{\alpha \beta}(r) = 4 \epsilon_{\alpha \beta} \left[ \left(
\frac{\sigma_{\alpha \beta}}{r} \right)^{12} - \left(
\frac{\sigma_{\alpha \beta}}{r} \right)^{6} \right]. 
\end{eqnarray} 
This system has been extensively studied by
Wahnstrom~\cite{Wahnstrom1991443752}, Schr\o der~\cite{tbs}, and our
group~\cite{Glotzer2000112509,Schroder20001129834}. Following previous work,
we use length parameters $\sigma_{A A}=1$,
$\sigma_{B B}=5/6$, and $\sigma_{A B} = (\sigma_{A A} + \sigma_{B
B})/2 $, and energy parameters $\epsilon_{A A}=\epsilon_{B B}
=\epsilon_{A B}=1$.  The masses of the particles are chosen to be
$m_{A}=2$ and $m_{B}=1$.  We shift the potential and truncate it so it
vanishes at $r=2.5\sigma_{A B}$.

We simulate a system of $N=8000$ particles using periodic boundary
conditions in a cubic box of length $L=18.334$ in units of $\sigma_{A
A}$, which yields a density of $\rho = N/L^3=1.296$ for all state
points.  We report time in units of $\tau = (m_{B} \sigma_{A
A}^2/48\epsilon_{AA})^{\frac{1}{2}}$, length in units of $\sigma_{A
A}$, and temperature, $T$, in units of $\epsilon_{AA}/k_{B}$, where
$k_{B}$ is Boltzmann's constant. We simulate state points in the $NVE$
ensemble at temperatures ranging from $T=2.0$ to $T=0.001$, following
a path similar to that followed in
Refs.~\cite{Glotzer2000112509,tbs,Schroder20001129834,Schroder1998235331}.
Additionally, we simulate a series of state points in the $NPT$
ensemble for temperatures ranging from $0.1$ to $0.6$ at zero average
pressure. Simulation details can be found in
Refs.~\cite{LacevicJCP2003,NLthesis}. We estimate the mode coupling
temperature $T_{{\rm MCT}}=0.57 \pm 0.01$, (the glass transition
temperature $T_g$ is typically in the range $0.6T_{MCT} < T_{g} <
0.9T_{MCT}$\cite{Novikov200367031507}) and the Kauzmann temperature
$T_{0}$, which can be considered a lower bound for the glass
transition temperature $T_{{\rm g}}$, is $T_0 = 0.48 \pm 0.02$. These
quantities are found by calculating the structural relaxation time
$\tau_{\alpha}$ by fitting the secondary relaxation of the coherent
intermediate scattering function $F(q_0,t)$, evaluated at wavevector
$q_0$ corresponding to the maximum peak in the static structure
factor, to a stretched exponential function $F(t) = A {\rm
exp}(-(t/\tau_{\alpha})^{\beta})$. Table~\ref{tab1} summarizes the
$T$-dependence of $\tau_{\alpha}$ for the simulations in the $NVE$
ensemble.
  
\begin{table}
\begin{tabular}{|c|c|}
\hline
 $\langle T \rangle$ & $\tau_{\alpha}$ 
\\ 
\hline
\colrule
   $0.588 \pm 0.001$ & $ 3500 \pm 100 $ 
\\ $0.598 \pm 0.002$ & $ 1900 \pm100 $ 
\\ $0.615 \pm 0.001$ & $ 880 \pm 50 $ 
\\ $0.637 \pm 0.001$ & $ 370 \pm 50 $ 
\\ $0.660 \pm 0.001$ & $ 240 \pm 30 $ 
\\ $0.689 \pm 0.001$ & $ 150 \pm 30 $ 
\\ $0.944 \pm 0.001$ & $ 16 \pm 5 $ 
\\ $2.004 \pm 0.001$ & $ 4 \pm 1 $ 
\\
\hline
\end{tabular} 
\caption{\label{tab1} Average temperature $\langle T \rangle$ and
relaxation time $\tau_{\alpha}$}
 
\end{table}

\section{Basic quantities used to measure spatially heterogeneous dynamics}
\label{summary_SHD_4pnt}

The quantities relevant to SHD that we use here were calculated in
Ref.~\cite{LacevicJCP2003} using a theoretical framework based on a
four-point, time-dependent, density correlation function $g_4(r,t)$.
Here, we give definitions of these quantities that will be used in
later sections of the paper.  The quantity $g_4(r,t)$ is related to an
order parameter $Q(t)$ corresponding to the number of ``overlapping''
particles in a time window $t$, where the term ``overlap'' is used to
denote a particle which was either localized or replaced in a time
$t$. Mathematically, $Q(t)$ is defined as
\begin{eqnarray}
  Q(t) = \int d\textbf{r}_1 \textbf{r}_2 \rho(\textbf{r}_1,0)
\rho(\textbf{r}_2,t) w(|\textbf{r}_1 - \textbf{r}_2|) = \sum_{\rm{i=1}}^{N}
\sum_{\rm{j=1}}^{N} w(|\textbf{r}_{i}(0) - \textbf{r}_j(t)|),
\label{eqn5}
\end{eqnarray}
where $w(|\textbf{r}_1 - \textbf{r}_2|)$ is unity
for $|\textbf{r}_1 -\textbf{r}_2| \leq a$ and zero otherwise. We
take $a=0.3$. The
reason for introducing an ``overlap'' function $w$ is to eliminate
the vibrational motion of the particles, which is known to be only
weakly correlated at best (for more
details  see e.g. Ref.~\cite{LacevicJCP2003}). In
Refs.~\cite{Glotzer2000112509,LacevicJCP2003}, we showed that $Q(t)$
can be decomposed into self and distinct components,
\begin{eqnarray}
 Q(t) &=& Q_S(t) + Q_D(t) = \sum_{\rm{i=1}}^{N} w(| \textbf{r}_{i}(t) -
\textbf{r}_{i}(0)|) \nonumber \\ &+& \sum_{\rm{i=1}}^{N} \sum_{\rm{i \not= j} }^{N} w(|
\textbf{r}_{i}(0) - \textbf{r}_{j}(t)|).
\end{eqnarray}
The self part, $Q_S(t)$, measures the
number of particles that move less than a distance $a$ in a time interval
$t$; we call these ``localized'' particles.
The distinct part, $Q_D(t)$ measures the
number of particles replaced within a radius $a$ by another particle
in time $t$; we call these ``replaced'' particles. 

We also consider ``delocalized'' particles, that is, particles that in
a time $t$ moved more than a distance $a$ from their original
location.  As was pointed out in Ref.~\cite{Glotzer2000112509},
substituting $1-w$ for $w$ in Eq.~(\ref{eqn5}) gives the delocalized
order parameter $Q_{DL}(t)= N - Q_{S}(t)$.

The fluctuations, $\chi_4(t)$, in $Q(t)$ may be defined as
\begin{eqnarray} 
\chi_4(t) = \frac{\beta V}{N^2}[\langle Q(t)^2 \rangle - \langle Q(t)
  \rangle^2].
\label{eqn4-1} 
\end{eqnarray} 
Following the scheme of decomposing $Q(t)$, $\chi_4(t)$ can be
decomposed into self $\chi_{SS}(t)$, distinct $\chi_{DD}(t)$, and cross
$\chi_{SD}(t)$ terms.  $\chi_{SS}(t) = \chi_{DL}(t)$
is the susceptibility arising from fluctuations in the number of
localized particles, $\chi_{DD}(t)$ is the susceptibility arising from
fluctuations in the number of particles that are replaced by a
neighboring particle, $\chi_{SD}(t)$ represents cross fluctuations
between the number of localized and replaced particles, and
$\chi_{DL}$ is the susceptibility arising from fluctuations in the
number of delocalized particles.

$\chi_4(t)$ (and its terms) measures the correlated motion between
pairs of particles, calculated equivalently from fluctuations in the
number of overlaps or from the four-point correlation function
itself. As shown in Ref.~\cite{LacevicJCP2003}, the behavior of
$\chi_4(t)$ demonstrates that correlations are time dependent, with a
maximum at a time $t_4^{\rm{max}}$. The characteristic times
$t_4^{\rm{max}}$ and $\tau_{\alpha}$ (defined in the previous section)
have similar $T$-dependence, indicating that the correlations measured
by $\chi_4(t)$ are most pronounced in the structural
$\alpha$-relaxation regime.

\section{Average force and instantaneous force distribution function $P(F)$} 
\label{AvgForInst}

\begin{figure}[tbp]
\begin{center}
\includegraphics[clip,width=15cm]{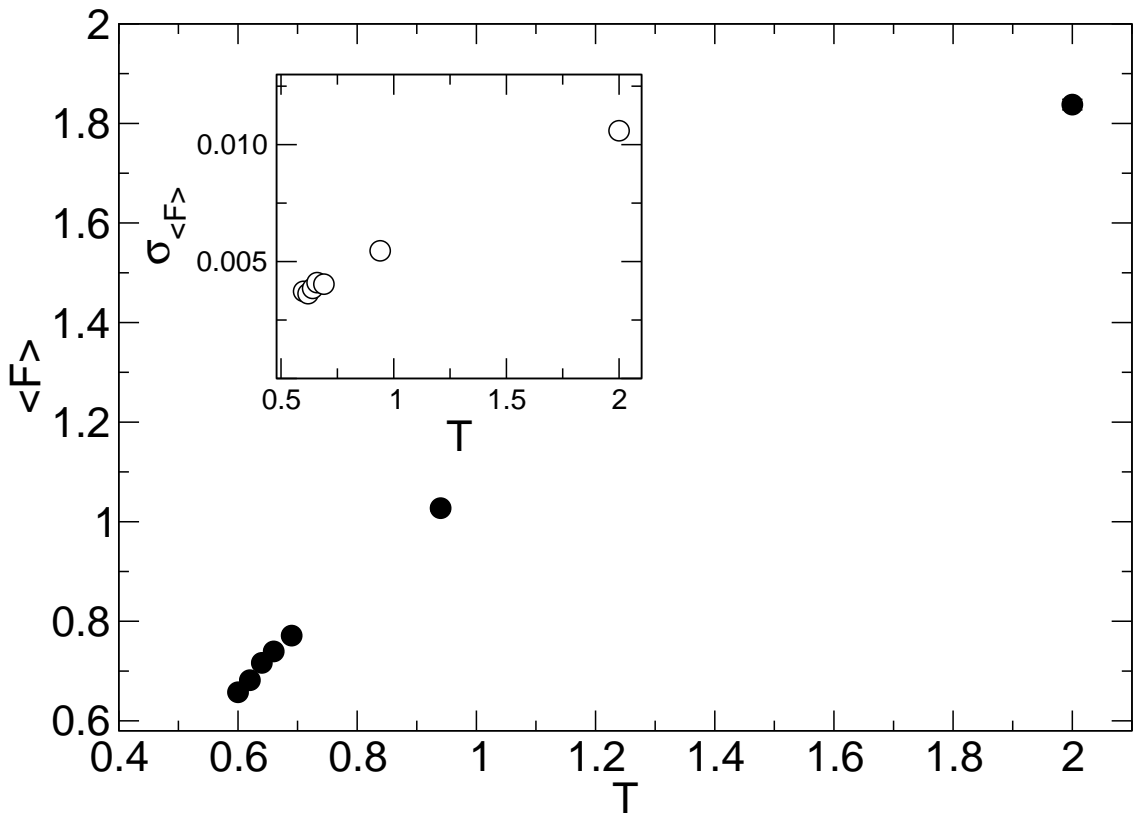}
\end{center}
\caption{$T$-dependence of the average force $\langle F \rangle$. The
inset shows the $T$-dependence of the standard deviation $\sigma_{\langle
F \rangle}$.  The error bars on $\langle F \rangle$ and
$\sigma_{\langle F \rangle}$ are smaller than the symbol size. }
\label{AverageForceSTD} 
\end{figure}

Because we are studying a dense liquid described by a relatively short
range pair potential, at an instant in time each particle experiences
a force due to the presence of neighboring particles. Following
Ref.~\cite{O'Hern200186111}, we calculate the average force $\langle F
\rangle$ between every pair of neighboring particles at a given
instant of time.  In Figure~\ref{AverageForceSTD} we show $\langle F
\rangle$ vs $T$ calculated from at least $N_{{\rm conf}} = 1000$
independent $NVE$ configurations at each $T$. We see that both $\langle F
\rangle$ and the standard deviation $\sigma_{\langle F \rangle}= \Big
(
\frac{1}{N_{{\rm conf}}(N_{{\rm conf}}-1)} \sum_{c} \Big (\langle F
\rangle_{c} - \langle F \rangle)^2 \Big )^{\frac{1}{2}}$, where
$\langle F \rangle_{c}$ is the average force for a particular
configuration ($c$), decrease with decreasing $T$. [We also find that
the standard deviation $\sigma_{\langle F \rangle}$ calculated within
$\it{each}$ configuration has the same temperature dependence (not
shown).] A decrease is expected since the average pressure (see
Table~1 in Ref~\cite{LacevicJCP2003}) in our constant volume
simulation decreases as $T$ decreases.  Large $\sigma_{\langle F
\rangle}$ at higher $T$ is also expected because of the larger
fluctuations in pressure at higher temperatures.  
$\langle F \rangle$ is calculated
in the following manner. First, the average force of a particular
configuration $c$ is found from $\langle F \rangle_{c} = (\sum_{ij}
F_{ij})/N_{ij}$, where $N_{ij}$ is the number of pairs $i$ and $j$
for which the force is nonzero, and particles $i$ and $j$ belong to that
particular configuration. An ensemble average force $\langle
F \rangle$ is calculated as an equally weighted average over all
$\langle F \rangle_{c}$. This can be an important point when there are
large force fluctuations that result in substantially different
numbers of positive forces for different configurations.  We have,
however, checked an alternative method to calculate average force by
calculating the average force from all particle pairs making no
distinction between configurations, and obtained the same answer,
indicating that the system is self-averaging. In
Ref.~\cite{O'Hern200288075507} the authors showed that the presence of
non-self-averaging alters the force distribution function $P(F)$,
which we now consider.

\begin{figure}[tbp]
\begin{center}
\includegraphics[clip,width=15cm]{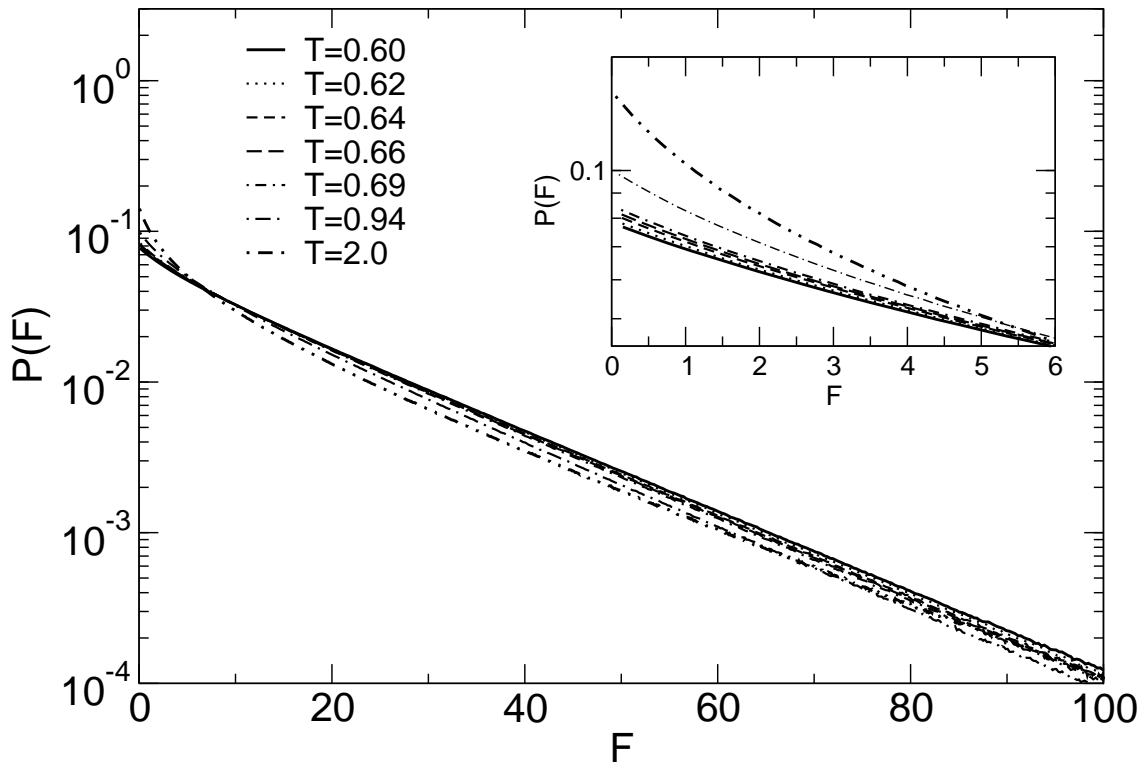}
\end{center} 
\caption{$T$ dependence of $P(F)$ vs $F$.  The inset shows behavior of
$P(F)$ for small positive forces. }  
\label{fig_ch5_1} 
\end{figure}

In foams and granular materials, $P(F)$ is measured as a distribution
of interparticle normal forces~\cite{Tewari1999604385,Langer20004968}.
In our system, there is no strict definition of point contact between
Lennard-Jones(LJ) particles, and therefore the normal force between
particles in not well-defined. We define two LJ particles to be in
contact if they interact with a positive (repulsive) force, as it was
done in Ref.~\cite{O'Hern200186111}. Commonly,
granular materials are modeled as systems of particles with only
repulsive interactions when grains are in contact. Therefore, we
investigate $P(F)$ only for the subset of positive (repulsive) $F$.
Figure~\ref{fig_ch5_1} shows the $T$-dependence of the force
distribution function $P(F)$, calculated as a normalized histogram of
all instantaneous positive forces between particle pairs.
The behavior of $P(F)$ is similar to the one observed in
Ref.\cite{O'Hern200186111} for their LJ system above $T_g$.  We
observe an exponential tail at high forces and an increase in the
curvature of $P(F)$ at small forces as $T$ decreases. An explanation
of the origin of the long exponential tail in $P(F)$ was proposed in
Ref.~\cite{O'Hern200186111} and is related to the fact that large $F$
behavior can be obtained from the small $r$ behavior of $g(r) \approx
y(r)\rm{exp}[-V(r)/k_bT]$, and $P(F)dF \propto g(r) dr$. The
exponential tail in the force distribution of granular materials was
suggested in the $q$-model in Ref.~\cite{Coppersmith1996534673} to be
a consequence of a force randomization throughout the packing. This
randomization has an effect analogous to the collisions in an ideal
gas~\cite{Liu1995269513,Coppersmith1996534673}.

\begin{figure}[tbp]
\begin{center}
\includegraphics[clip,width=15cm]{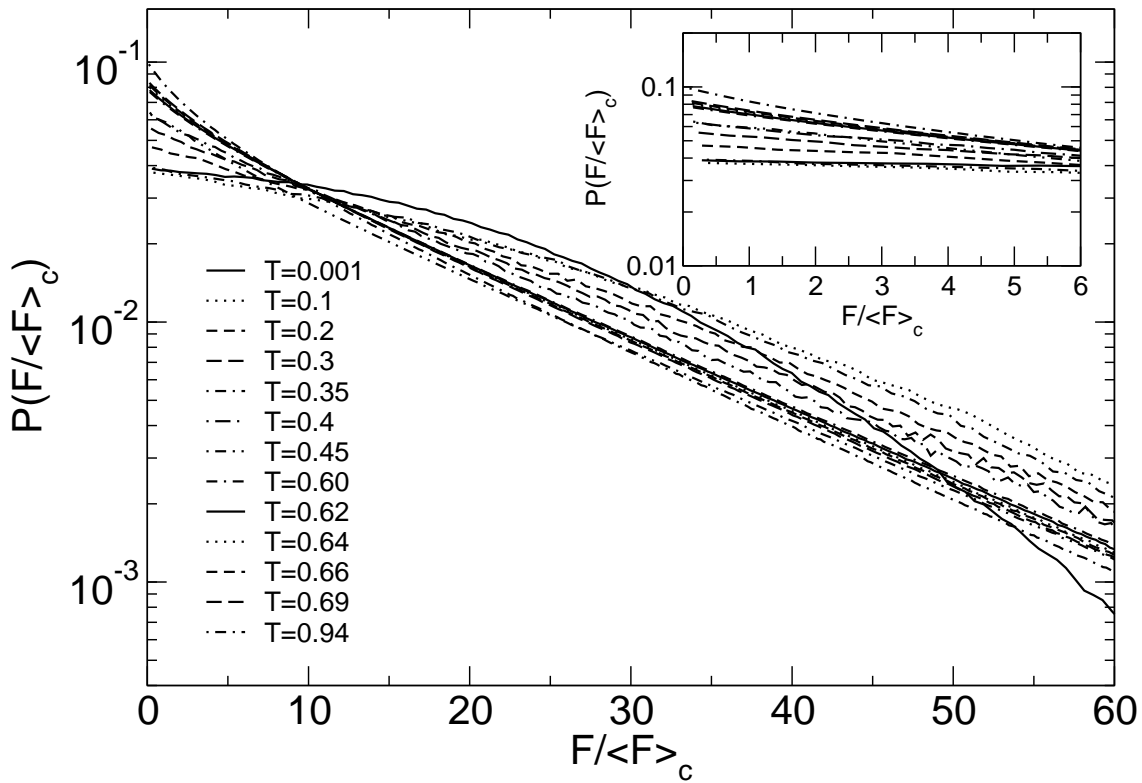}
\end{center}
\caption{$T$-dependence of the force distribution $\rm{P(F/\langle F \rangle_c)}$
vs $F/\langle F \rangle_c$. Forces that are much larger than $\langle F
\rangle$ decay exponentially. The curvature of $P(F/\langle F
\rangle_c)$ around $F/\langle F \rangle_c =1 $ does not change sign but
its value decreases as $T$ decreases. The inset shows a
close-up of $P(F)/\langle F \rangle_c$ for forces near $\langle F
\rangle_c$.}
\label{fig_ch5_2}
\end{figure}

To further compare $P(F)$ with that in granular materials, we
calculate the force distribution function scaled with the average
force $\langle F \rangle_{c}$ for a particular configuration,
$P(F/\langle F \rangle_{c})$.  The $T$ dependence of $P(F/\langle F
\rangle_{c})$ vs $F/\langle F \rangle_{c}$ is shown in Figure~\ref{fig_ch5_2}.
We include $P(F/\langle F \rangle_{c})$ for
several temperatures below $T_0 = 0.48$ for 
comparison with the force distribution found in Ref.~\cite{O'Hern200186111} 
below $T_g$. As in Figure~\ref{fig_ch5_1}, the probability of observing
forces substantially larger than $\langle F \rangle$ decays
exponentially, and the curvature around $\langle F \rangle$ decreases
as $T$ decreases (See Figure~\ref{fig_ch5_2} caption).  The resemblance of Figure~\ref{fig_ch5_1} and
Figure~\ref{fig_ch5_2} to similar figures in Ref.~\cite{O'Hern200186111}
again shows there is no significant distinction
between averaging pair forces within a configuration and globally
(across configurations), confirming that large force fluctuations,
like those found in Ref.~\cite{O'Hern200288075507} for
an out-of-equilibrium system, are not present here.

We note that the reason for using two methods for the calculation of
$P(F)$ is the fact that glass-forming liquids, such as our model, are
not in equilibrium on all time scales. Long lived
clusters of immobile particles have been reported to persist for 
times that are orders
of magnitude longer that the structural relaxation time
$\tau_{\alpha}$\cite{HeuerJNCS}. It is intuitive that if those
clusters exist for a long time, forces between mobile particles could
have large fluctuations in order to initiate structural
relaxation. Therefore, it is possible that large force fluctuations
will be present in supercooled liquids even at structural relaxation
times.  If this was the case, then one could observe non 
self-averaging of $P(F)$.  We demonstrate that this is not the case by
comparing two different methods of calculating $P(F)$.

In the experiments and simulations of granular materials and foams, it
has been suggested that a signature of jamming is the development of a
small peak or plateau around the average force. O'Hern et
al. suggested that $P(F)$ will develop a peak if the first peak of
$g(r)$ is ``sufficiently high and narrow''~\cite{OhernPRE} which they
associate with solid-like behavior of the system and a possible
signature of the glass transition.  Indeed we observe the sharpening
of the first peak in $g(r)$ (not shown here, see
e. g. Ref.~\cite{Lacevic200266030101}) in our system as $T$ decreases,
but the peak in $P(F)$ is absent for $T < T_0$.  Recall that the
jamming transition in granular materials is equivalent to the
macroscopic structural arrest of the system. This would correspond to
a glass transition in supercooled liquids. Thus for $T>T_g$, we would
not expect to observe the peak or plateau in $P(F/\langle F
\rangle_{c})$ (see Figure~\ref{fig_ch5_2}). In order to investigate
the behavior of $P(F/\langle F \rangle_{c})$ in the glass, we
calculate $P(F/\langle F \rangle_{c})$ at several $T$ below
$T_0$. These state points are obtained by quenching a configuration
previously equilibrated at high temperature, in our case $T=10.0$, to
a desired $T$, at a non-zero pressure.  We do observe that the
curvature of $P(F/\langle F \rangle_{c})$ around $\langle F \rangle$
decreases as $T$ decreases, and that the slope of $log P(F/\langle F
\rangle_{c})$ for $10 < \langle F \rangle < 60$ changes significantly
for $T=0.001$.  We observe that the peak or plateau is absent in our
system approaching $T_{\rm{MCT}}$ from above, and that the plateau
develops at much lower temperatures ($T=0.1$) compared to $T_0$.  This
demonstrates that development of a peak or plateau is not a necessary
condition for a glass transition or a solid-like behavior in dense
glass-forming systems.

\begin{figure}[tbp]
\begin{center}
\includegraphics[clip,width=15cm]{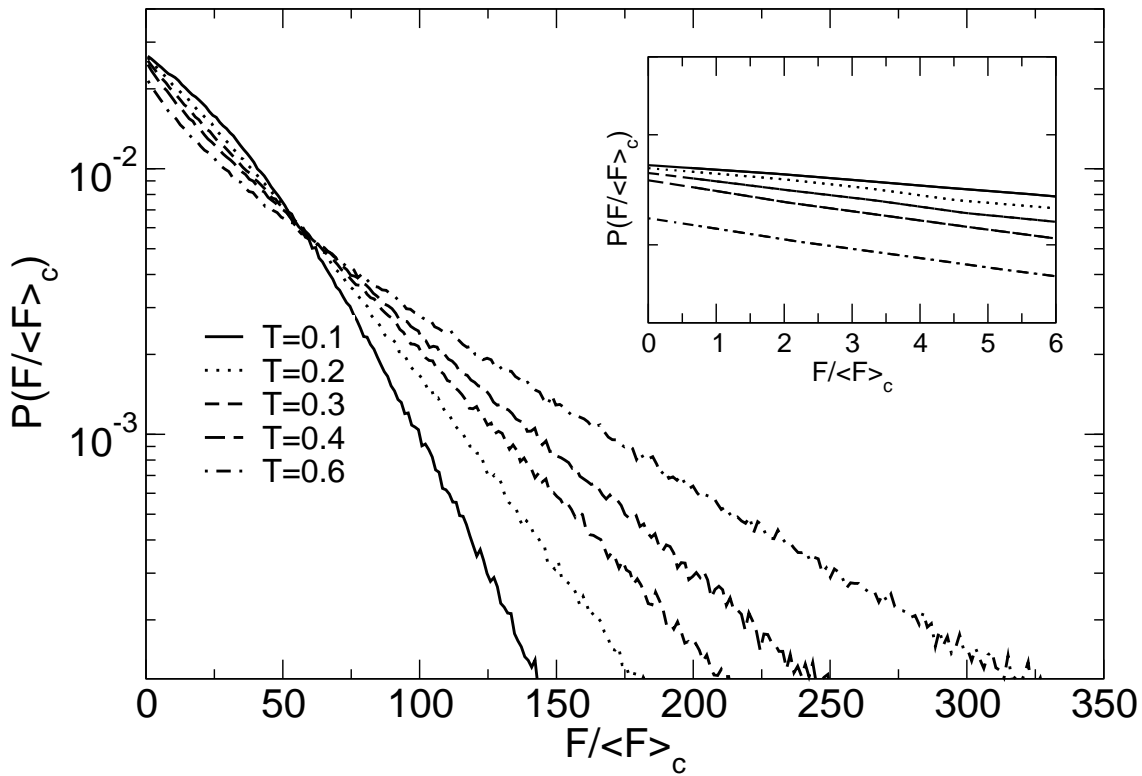}
\end{center} 
\caption{$T$-dependence of the force distribution $\rm{P(F/\langle F
\rangle_c)}$ vs $F/\langle F \rangle_c$ at several state points in
the $NPT$ ensemble at $P=0$. The inset shows a close-up
of $P(F)/\langle F \rangle_c$ for forces near $\langle F \rangle_c$.}
\label{NPT_forcedist-1} 
\end{figure}

In the simulations of Ref.~\cite{O'Hern200186111}, a barostat was used
to maintain an average zero pressure. To test if the peak development
in $P(F/\langle F
\rangle_{c})$ depends on the choice of thermodynamic ensemble, we
performed simulations of our $3D$ binary LJ system at several
temperatures in the $NPT$ ensemble, and calculated the corresponding
$P(F/\langle F \rangle_{c})$. Figure~\ref{NPT_forcedist-1} shows the 
$T$-dependence of $P(F/\langle F \rangle_{c})$ for several $NPT$ state
points.  The pressure is kept at zero for all temperatures.  We 
again observe a change in the curvature in $P(F)$, but only at very low
temperatures below $T_g$, consistent with what our findings for the 
$NVE$ system.  This again shows that a peak in $P(F/\langle F
\rangle_{c})$ is not necessary for solid-like behavior of our system.
Similar findings have been reported by Reichman and Sastry for their
model supercooled liquid~\cite{rs_privtcmm}.

It is interesting to note that both the $NVE$ and $NPT$ force
distributions $P(F/\langle F \rangle_{c})$ have a value of $F/\langle
F
\rangle_{c} $ where the distributions cross
for almost all $T$, indicating an isosbestic point. This value is at approximately $F/\langle F \rangle_{c}= 10$ and $55$ for the $NVE$ and $NPT$ systems,
respectively. The reason for such behavior is not obvious, and will be
investigated in future work. Also note that we tested the
self-averaging of $P(F/\langle F \rangle_{c})$ for all state points
for which the system is a glass. We find no evidence for non self
averaging of $P(F/\langle F \rangle_{c})$ at those points.

Despite the fact that the peak or plateau in $P(F)$ is observed in
many granular materials and some supercooled liquids below $T_g$,
an open question is its connection to the development of the yield
stress in those materials. Several speculations have been made in
Refs.~\cite{O'Hern200186111,OhernPRE} in order to explain the jamming scenario
and its possible connection to the glass transition.  They theorize that
``systems jam when there are enough particles in a force chain network
to support stress over the time scale of the measurement'' which would
imply that force chains observed in granular packing may also be
important to the glass transition. Further, since force chains do not
couple strongly to density fluctuations they may be linked to
local dynamical heterogeneities near $T_g$.
To test this speculation we seek to find a link between SHD and
particles that belong to the different regions of the force
distribution function, namely the particle pairs that interact with
forces that belong to the exponential tail and average force in the
force distribution function.

\section{Subsets of instantaneous forces}
\label{sub_force_inst}

\begin{figure}[tbp] 
\begin{center}
\includegraphics[clip,width=15cm]{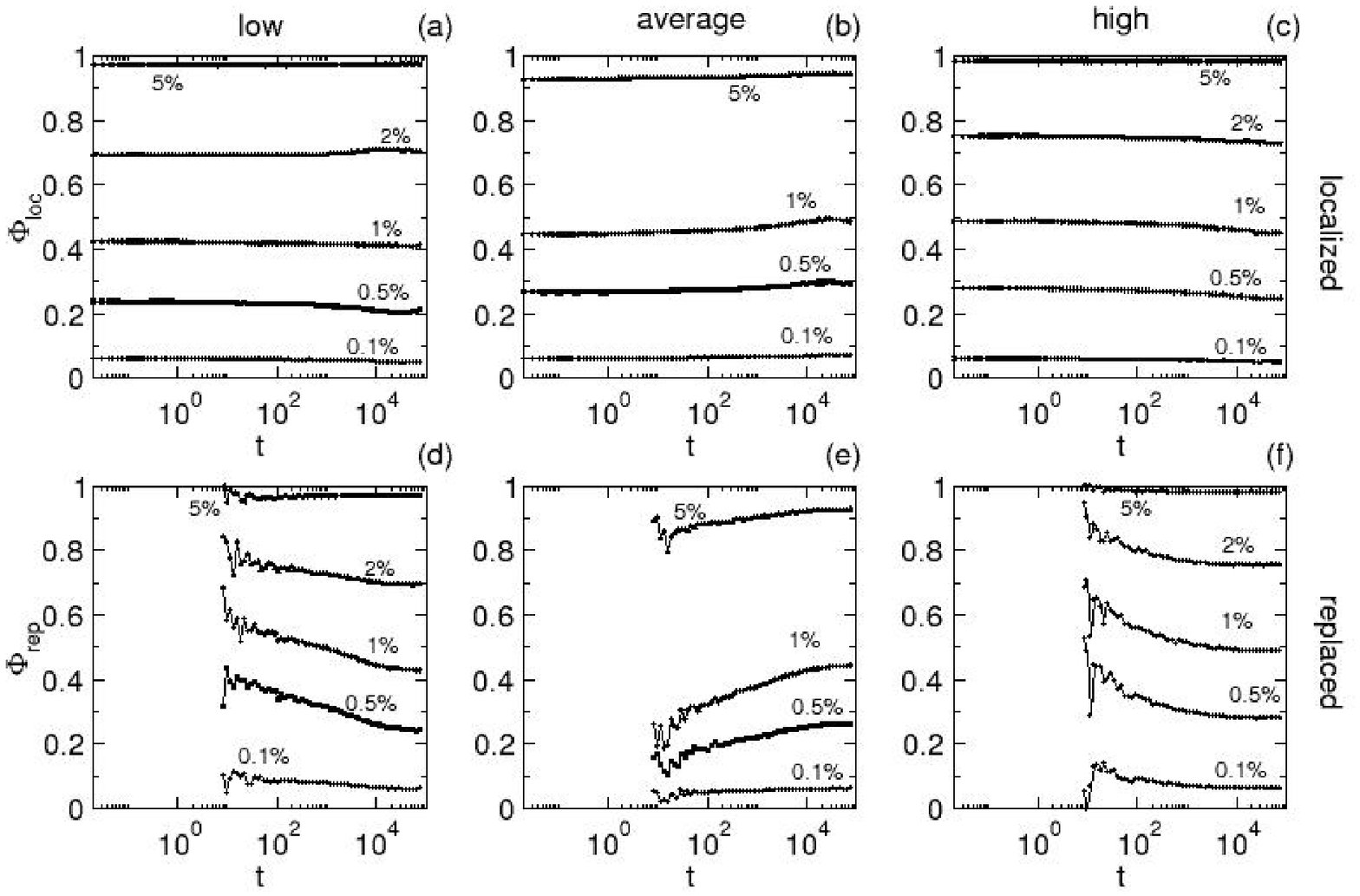}
\end{center}
\caption{Time dependence of $\Phi_{{\rm loc}}$ (upper three panels) and
$\Phi_{{\rm rep}}$ (lower three panels) in the subsets of low, average and
high pair forces at $T=0.60$. $\Phi_{{\rm rep}}$
is shown for times when $Q_D$ becomes nonzero. Labels on the $y$ axes
for average and high forces are omitted because of the clarity.}
\label{fig_ch5_3} 
\end{figure}

In Ref~\cite{LacevicJCP2003}, we established a criterion to find
localized and replaced particles which we use here to test the
relationship between particle mobilities and instantaneous pair
forces. A brief description of how to identify localized, replaced and
delocalized particles can be found in Section~\ref{summary_SHD_4pnt}.
For every two configurations (one at $t=0$ and the other at $t$) we find the number
of localized ($Q_S(t)$) and replaced ($Q_D(t)$) particles and
calculate all instantaneous pair forces in the configuration at time
$t$. We sort all instantaneous pair forces according to their values
and find subsets of particles that interact with pair forces that fall
within certain percentages of the lowest, average and highest pair
forces.  The considered percentages of these forces are indicated in
Figure~\ref{fig_ch5_3}.
 
We define a fraction, $\Phi_{{\rm loc}}$ ($\Phi_{{\rm rep}}$), as the ratio of the
number of localized (replaced) particles that are associated with the
given subset of pair forces to the total number of localized
(replaced) particles, $Q_S$ ($Q_D$). $\Phi_{{\rm loc}}$ and $\Phi_{{\rm rep}}$
thus relate mobility and instantaneous forces.  Figure~\ref{fig_ch5_3}
shows the time dependence of $\Phi_{{\rm loc}}$ and $\Phi_{{\rm rep}}$ in the
given subsets of pair forces at $T=0.60$.  

For the limiting case where all localized and replaced particles
belong to the given subset of pair forces, $\Phi_{{\rm loc}} \equiv 1$ and
$\Phi_{{\rm rep}} \equiv 1$. This is true for subsets larger than $\approx
5\%$ of the pair forces. These fractions are large enough to include
nearly all of the system's particles since there are
many more pair forces ($ \approx 225000$) than particles ($8000$). For lower
percentages of the highest pair forces, $\Phi_{{\rm loc}}$ decreases in time
from the value expected for the bulk (at t=0 when all particles are
localized and $\Phi_{{\rm loc}}$ is simply the probability of finding any
particle in the given percentage of pair forces).  This suggests that because
particles that have been localized for long times are less likely to
experience high relative forces, if the reason for their long
localization is due to being at a local energy minimum.  $\Phi_{{\rm rep}}$
decreases in time toward the bulk value (at later times), but this
decrease starts on much shorter time scales than $\Phi_{{\rm loc}}$, possibly
because particles that have been replaced (mobile particles)
experience higher relative forces at times when they escape from their
cages.  In the case of low pair forces, the results for $\Phi_{{\rm loc}}$
and $\Phi_{{\rm rep}}$ should be taken with caution because there are two
ranges of distance at which particles can experience the lowest force
(the tail and well of the potential).  The ambiguity in low forces is
seen in the inconsistent behavior of $\Phi_{{\rm loc}}$ and $\Phi_{{\rm rep}}$.
This may mask any clear interpretation of the meaning of $\Phi_{{\rm loc}}$
and $\Phi_{{\rm rep}}$ for the lowest forces.  In the case of average pair
forces, the results are reversed from those of the high pair
forces. This means that a larger fraction of localized particles tend
experience average forces at later times, and replaced particles
tend to approach the bulk value of the average force.

This connection between mobility and instantaneous forces
leads us to investigate the spatial correlations of pairwise forces in
supercooled liquids.

\section{Generalized force chains in supercooled liquids}
\label{force_chains}

We now investigate the spatial correlations among the instantaneous pairwise 
forces. Spatially correlated forces can be viewed as
networked ``paths'' of highest, average or lowest forces in the system. 
Highest-force paths would correspond to the force chains in 
granular materials, and we refer to them as force chains.

According to Ref.~\cite{Cates1998811841}, a force chain in a granular
material is a ``linear string of rigid particles in point contact''. A
force chain in a granular material can support a load along its
axis. Since we do not apply a load or stress to our system, if such
structures exist, they would be induced by lowering the
temperature. Inspired by the work of Makse et
al.~\cite{Makse2000844160}, we look for these chains by searching the
``paths'' that may be found following instantaneous forces. In
Ref.~\cite{Makse2000844160}, the authors found force chains by starting
from a sphere at the top of the box of simulated spherical grains and
following the path of maximum contact force at every grain. They
observed a force-bearing network that was concentrated in a few
percolating chains.

\begin{figure}[tbp]
\begin{center}
\includegraphics[clip,width=15cm]{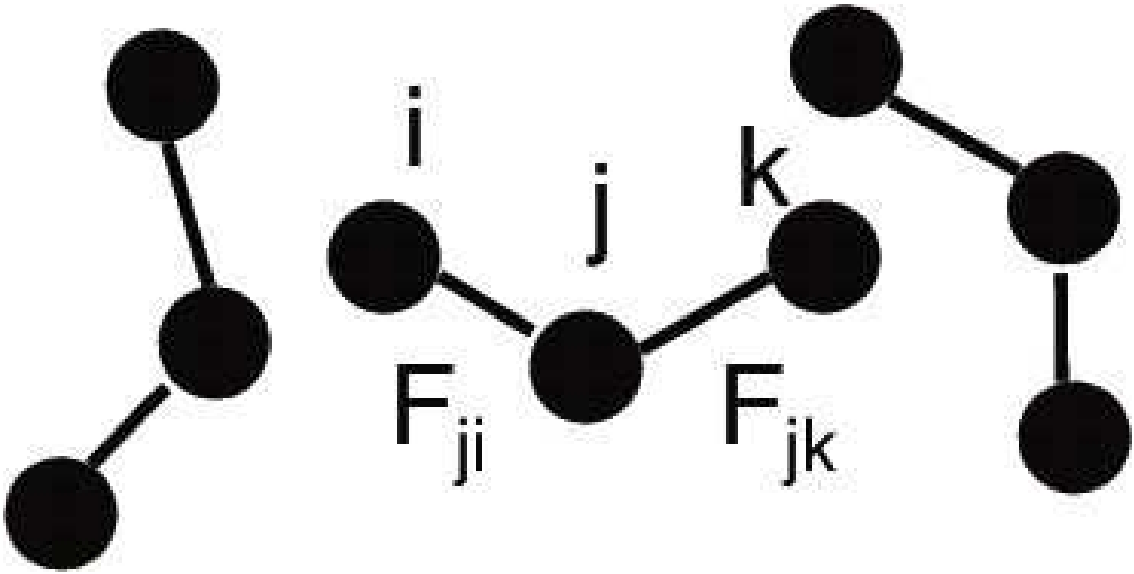}
\end{center}
\caption{Schematic picture of trivial force chains. Each particle
is connected to two neighbors with which it interacts with the highest
forces.  $F_{ji}$ and $F_{jk}$ are the largest forces on particle
$j$.  A similar picture can be drawn for interactions with low or
``closest to average'' forces.}
\label{Fig4.6}
\end{figure}

\begin{figure}[tbp]
\begin{center}
\includegraphics[clip,width=15cm]{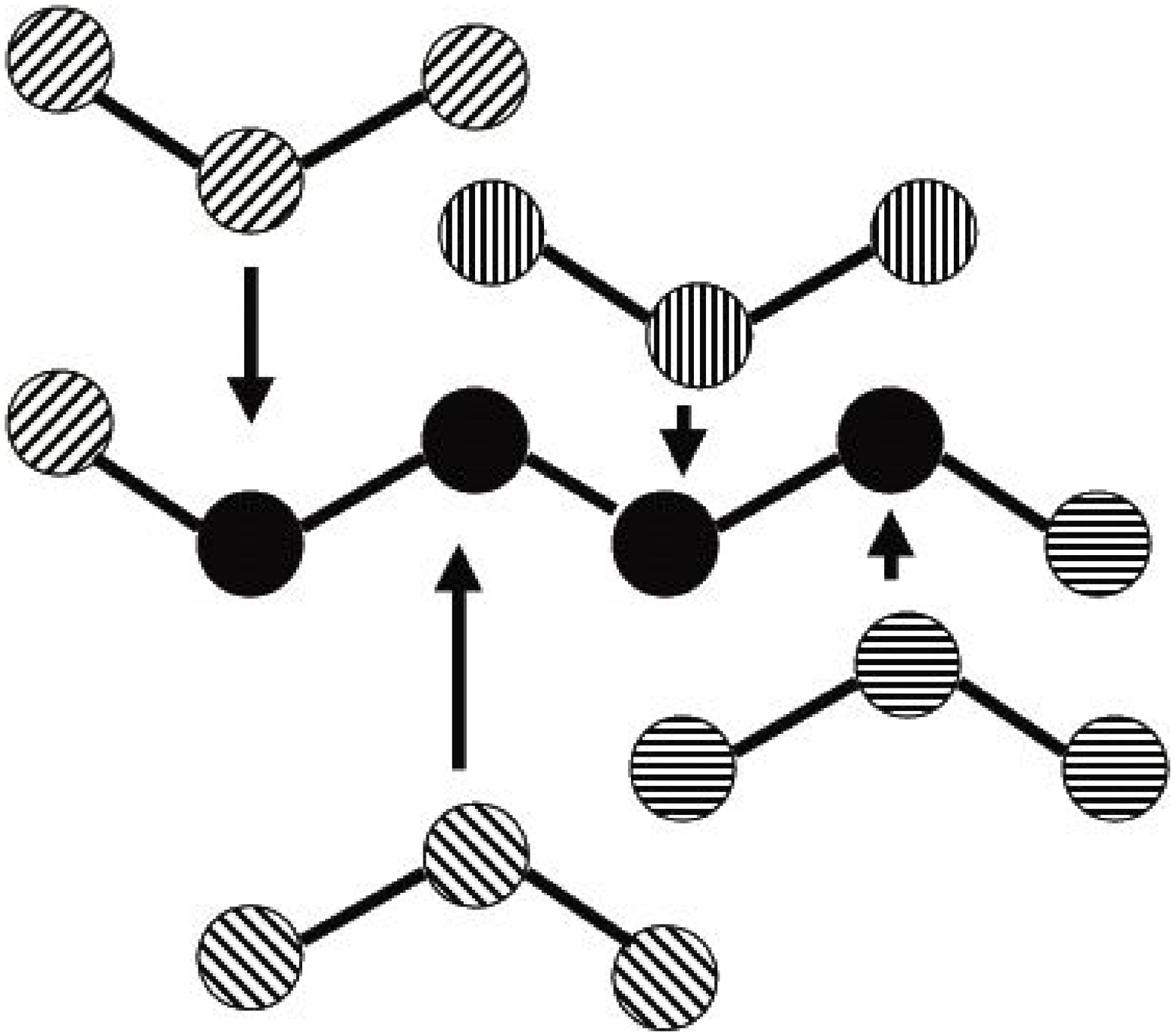}
\end{center}
\caption{Schematic picture of formation of non-trivial chains.
This is an example of a chain of mass six. In this example four trivial
chains combine to form a chain of mass six.}
\label{Fig4.7}
\end{figure}

In our case, we define force chains such that, for each particle, we
find the two neighboring particles that exert on that particle
the highest force.  This set of three particles constitute
what we call a ``trivial chain'', since it is always formed by construction.
These trivial chains are schematically shown in Fig.~\ref{Fig4.6}. 
Later, chains that are defined using low and average forces are also 
calculated in a similar manner as chains defined using the highest forces.
Longer chains are formed from trivial chains that share 
two members (Figure~\ref{Fig4.7}).
Force chains from the highest forces would correspond to force chains of point
contacts in granular materials.  We investigate low and average force
chains for the sake of completeness and the fact that we do not have a
theory that would a priori suggest whether jamming occurs because 
of high forces or,
e. g., average forces.

\begin{figure}[tbp]
\begin{center}
\includegraphics[clip,width=15cm]{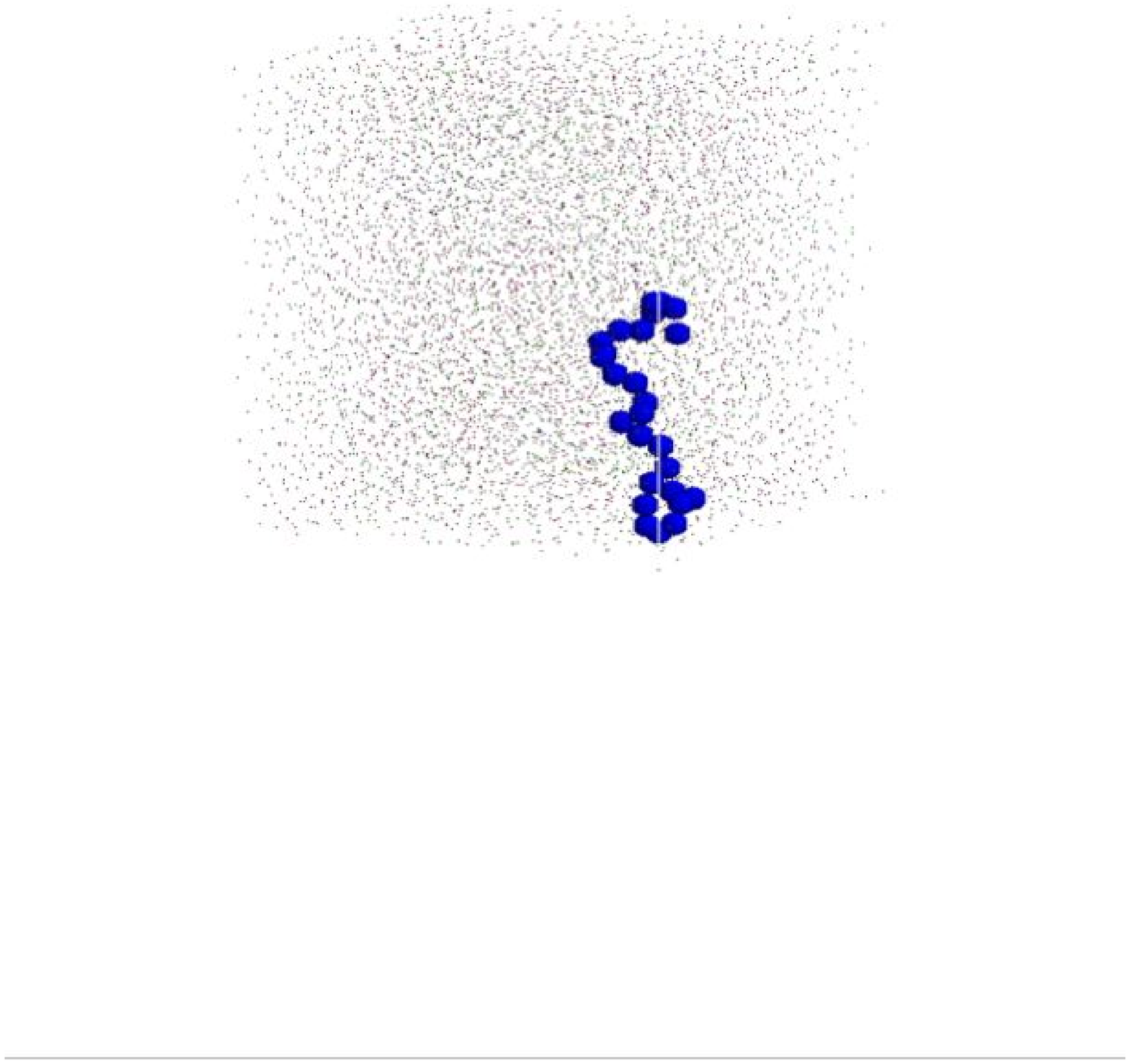}
\end{center}
\caption{$26$-particle chain at $T=0.60$. The particles in the chain
are shown with radius $0.5\sigma$. This is the longest chain for this
particular snapshot. This chain does not span the box.
Dots represent the rest of the $8000$ particles in the system.}
\label{chains26examp}
\end{figure}

\begin{figure}[tbp]
\begin{center}
\includegraphics[clip,width=15cm]{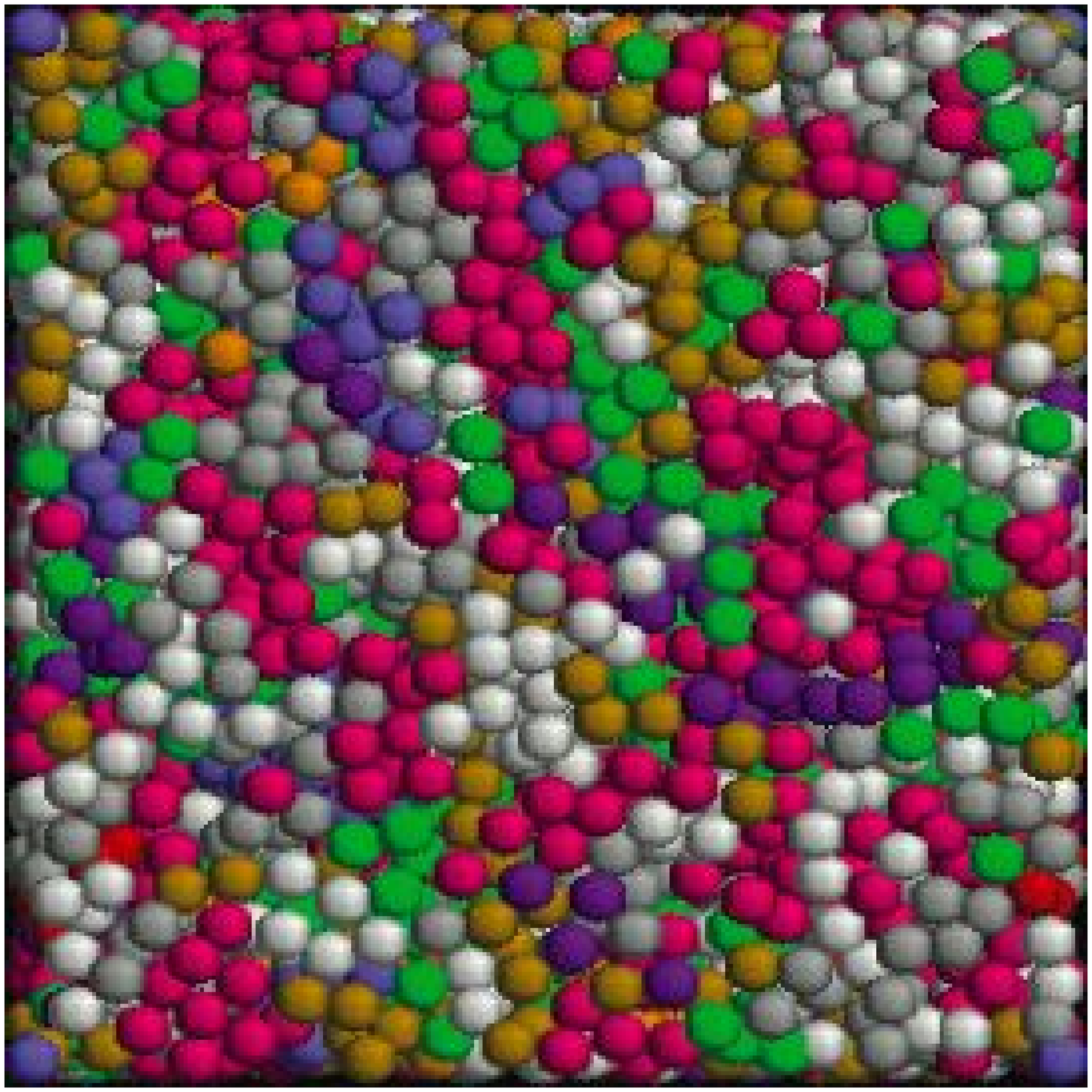}
\end{center}
\caption{Snapshot of all force chains at T=0.60. Different colors correspond to
force chains with different masses.}
\label{chainsT60examp}
\end{figure}

Figure~\ref{chains26examp} shows a chain of highest forces containing
$26$ particles in a single configuration at $T=0.60$. Figure~\ref{chainsT60examp}
shows all highest force chains at $T=0.60$ at one snapshot.
We refer to chain mass $n$ as the number of particles in that
chain. $N_{\rm chains}$ represents the number of non-trivial chains for a
particular configuration. 

\begin{figure}[tbp]
\begin{center}
\includegraphics[clip,width=15cm]{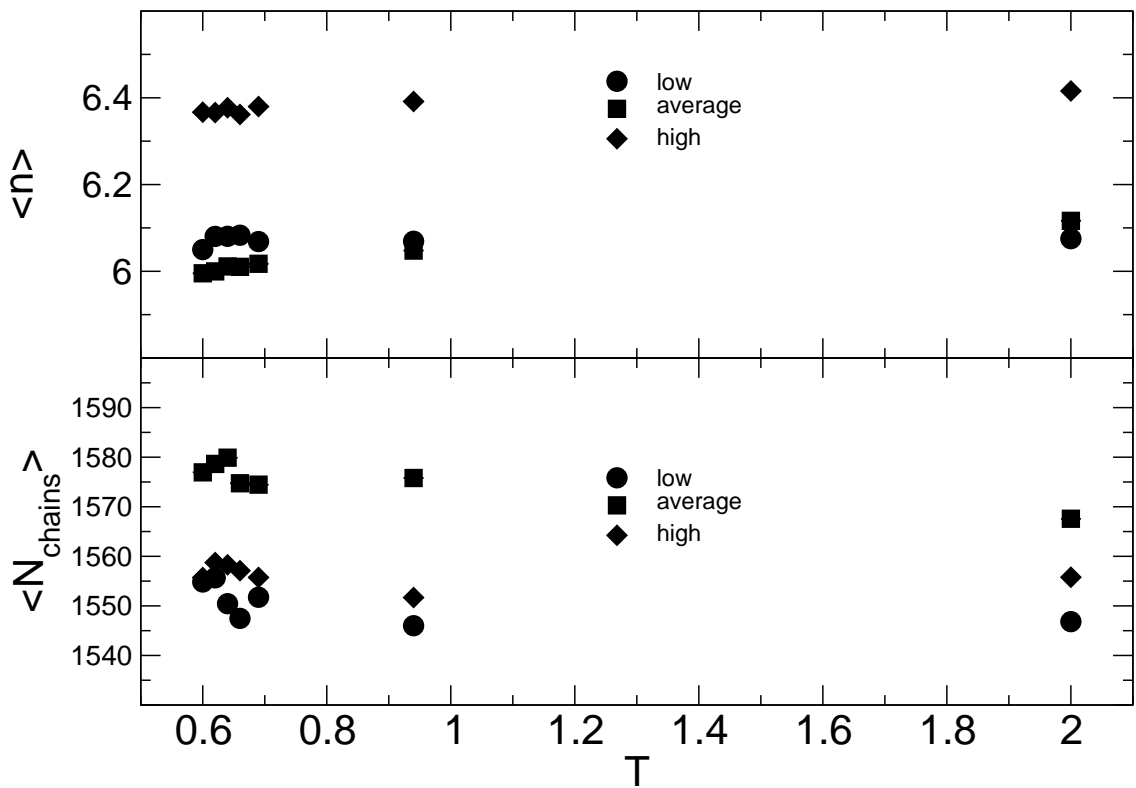}
\end{center}
\caption{$T$ dependence of the (a) average mass $n$ of the force chains and (b)
average number $\langle N_{\rm chains} \rangle$ of non-trivial force chains for low, 
average and high pair forces. The error bars are smaller than the symbol size.}
\label{AvMassAvNochains}
\end{figure}

The first trend we investigate is the $T$-dependence of the average
mass $\langle n \rangle$ of the force chains and the average number of
non-trivial chains $\langle N_{\rm chains} \rangle$. These quantities
are shown in Figure~\ref{AvMassAvNochains}.
$\langle n \rangle$ and $\langle N_{\rm chains} \rangle$ are averages
over the same configurations we used to calculate the force
distribution functions, $P(F)$, in Section~\ref{AvgForInst}.  It is
evident from Figure~\ref{AvMassAvNochains} that $\langle n \rangle$
and $\langle N_{\rm chains}
\rangle$ have at most a very weak temperature dependence.
We observe that chains with high forces are on average slightly longer
than chains with average forces.  We also observe that the average
number of chains calculated for the average force subsets is slightly
larger than the average number of chains calculated for low and high
forces.

In order to look for any spanning chains (similar to
Ref.~\cite{Makse2000844160}), we calculate the radius of gyration and
end-to-end distance of the force chains, common quantities
used to describe the size and shape of polymers~\cite{Doi1988xiii}. We
find that these quantities are also largely independent of temperature
(not shown). Therefore, the geometry of force chains does not change
significantly with $T$. We also do not find any spanning chains that
would correspond to the case where the end-to-end distance is equal to
the size of the box or where the radius of gyration equals half of
the box size.  

\begin{figure}[tbp]
\begin{center}
\includegraphics[clip,width=15cm]{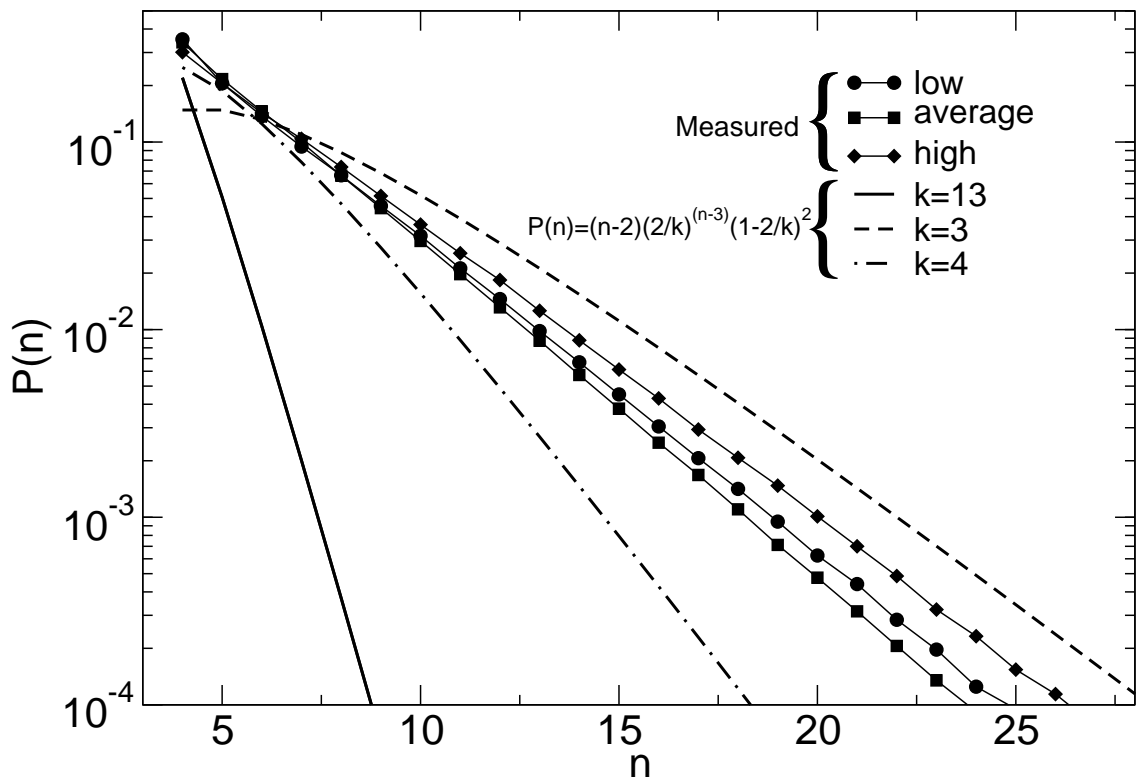}
\end{center}
\caption{Comparison of the measured mass distributions of force chains 
for low, average and high force at $T=0.60$ and theoretical predictions 
for different nearest neighbor parameter, $k$, values.}
\label{T60_fcdist}
\end{figure}

In order to better understand the difference between chains that are
defined using different subsets of forces, we investigate the
distribution function of chain mass $P(n)$ for chains defined using
low, average and high forces.  $P(n)$ is the probability of finding
chains of mass $n$ using the given subset of forces. $P(n)$ is calculated
as a normalized histogram of chain masses for each configuration and
averaged over all configurations. Figure~\ref{T60_fcdist} shows $P(n)$
vs $n$ for chains defined for each subset. We see from $P(n)$ that
chains of high forces are, on average, longer than chains of average
and low forces, as seen in Figure~\ref{AvMassAvNochains}(a).  Shorter
chains defined using average forces may be the consequence
of more branching of these chains compared to the chains composed of
the highest and lowest forces, which is consistent with the fact that
the number of force chains for the average forces is the largest.

\begin{figure}[tbp]
\begin{center}
\includegraphics[clip,width=15cm]{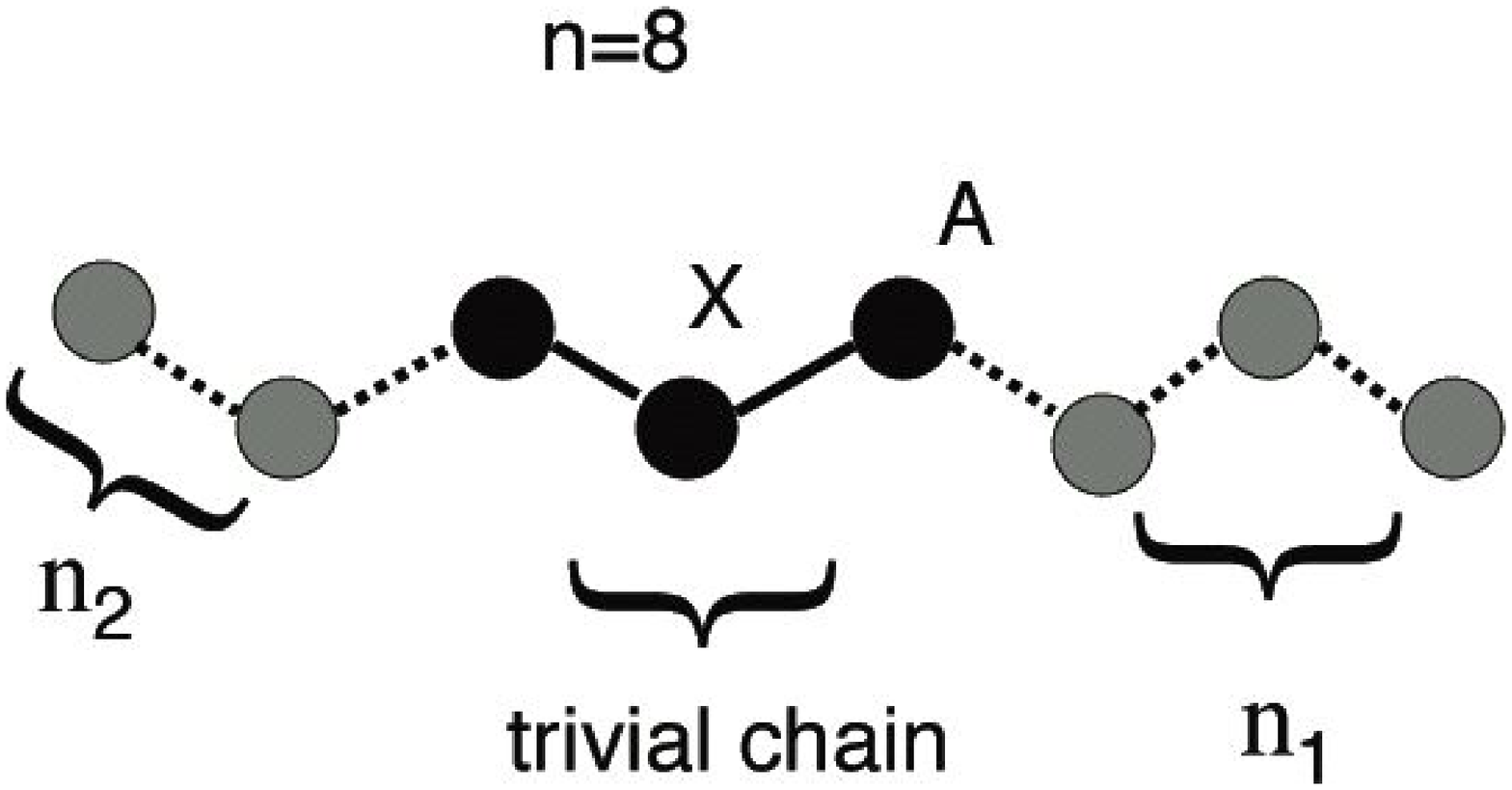}
\end{center}
\caption{Example of chain length prediction.}
\label{figExCh}
\end{figure}

To give a theoretical prediction of $P(n)$ for comparison, consider
the case where the forces between particles are randomly
assigned. Assume that there are $k$ nearest neighbors for a single
particle $X$. If a particle, $X$, shares one of its highest pair forces
with a neighboring particle, $A$, the probability that the interaction
between $A$ and $X$ is one of $A$'s two highest pairwise forces can be
approximated as $\Big ( \frac{2}{k} \Big )$.  Assuming that the chain
continues to grow away from its origin and the likelihood of it
looping back on itself is small, the probability of adding $n_1$
additional members on one end of a trivial chain ($n=3$)
(Figure~\ref{figExCh}) is
\begin{eqnarray} P(n_1) =
\Big (\frac{2}{k} \Big )^{n_1} \Big (1-\frac{2}{k} \Big ). 
\end{eqnarray} Since there are
two ends to the chain and the trivial chain always has three members,
the probability for a total chain length $n=n_1 + n_2 + 3$ is
\begin{eqnarray} 
P(n)= \Big (n-2 \Big) \Big ( \frac{2}{k} \Big )^{n-3}\Big (1-\frac{2}{k} \Big)^2,
\end{eqnarray} 
where the linear factor $n-2$ accounts for all
combinations of $n_1$ and $n_2$ that sum to $n-3$.
Figure~\ref{T60_fcdist} shows predictions of $P(n)$ for various values
of the nearest neighbor number parameter, $k$.  From the integration
of $g(r)$, the actual number of neighbors in the first neighbor shell
is approximately $13$.  As seen from the figure, the measured $P(n)$ decays
much more slowly than the prediction for entirely randomly assigned
forces.  Instead, the observations are well bounded by $k$ values of
three and four.  This gives some measure of the variability in the
local force distribution.  In other words, the more the local
environment of one particle varies from that of its neighbor, the
larger the value of $k$ that is found. We also note the possible
connection between the force chains found here and the idea of
rigidity percolation by Phillips~\cite{phillips} and
Thorpe~\cite{thorpe}. In rigidity percolation, the mean coordination number 
$m=2.4$ represents a threshold below which a network
glass is easily deformed. The fact that our nearest neighbor parameter $k$
is close to three suggests the possibility that the network of bonds in 
network glasses and the network of force chains in the LJ mixture studied 
here share similar properties, and one could use ideas developed
in rigidity percolation theory to study force chains in supercooled, 
non network-forming liquids.

\begin{figure}[tbp]
\begin{center}
\includegraphics[clip,width=15cm]{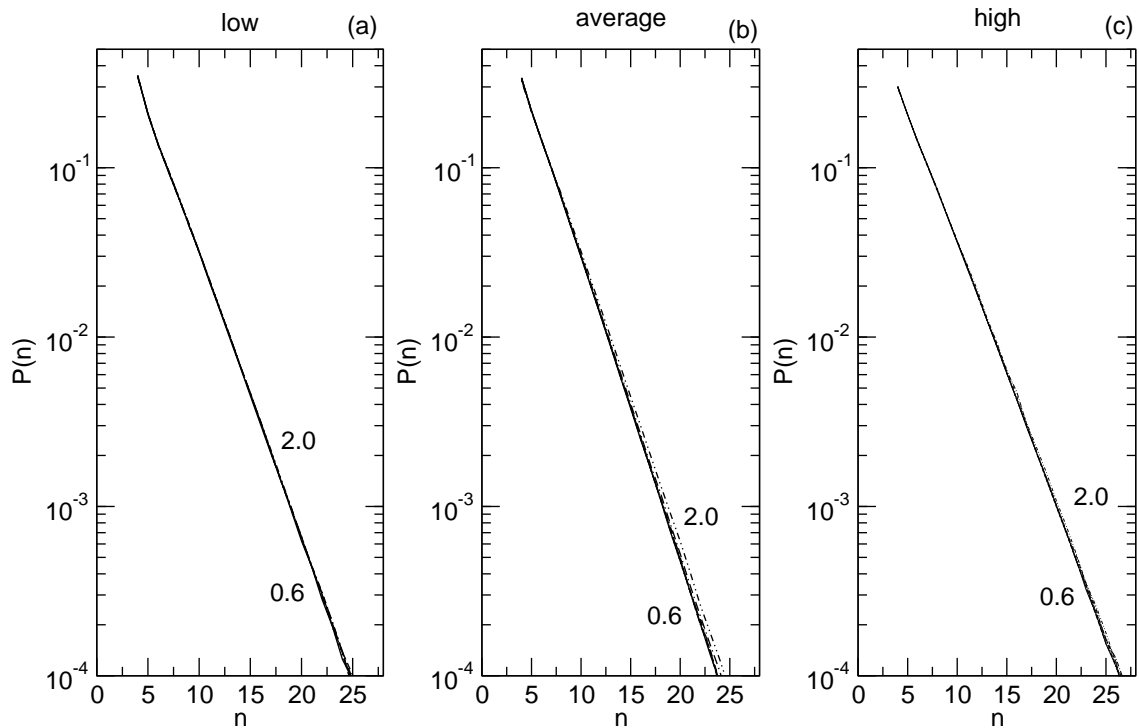}
\end{center}
\caption{$T$ dependence of the chain mass distribution $P(n)$ for the 
(a) low, (b) average and (c) high force. Force chains defined using
the average force subset show a slight $T$-dependence.}
\label{FCDist}
\end{figure}

Figure~\ref{FCDist} shows the $T$-dependence of $P(n)$ for chains in
low, average and high force subsets. We find that $P(n)$ can be
generally fitted well by a functional form $P(n) =
a_{1}\rm{exp}[-a_{2} n]$, where $a_1$ and $a_2$ are fitting
parameters. We note that observed exponential behavior of $P(n)$ is
analogous to that reported for equilibrium
polymerization~\cite{polymerization} of linear polymer chains, in
which the bonds between monomers break and recombine at random points
along the backbone of the chain. This picture is exactly what happens
to the force chain network in our system. Furthermore, the size
distribution of strings of cooperative rearranging particles is
exponential in all studies of strings performed thus far. In the case
of chains defined using average forces we note a slight $T$-dependence
of $P(n)$. It appears that these chains are shorter at lower
temperatures, but the distribution is still exponential. In the case
of the chains defined using the smallest forces, chains with mass
$n=4$ and $n=5$ do not fit well with the exponential function (note
the slight bend in the curve in Figure~\ref{FCDist}(a)).

We note that the force chains, as constructed, may be highly
susceptible to thermal fluctuations. More subtle trends may be
detectable by using methods to minimize these fluctuations such as
quenching to the inherent structure or time averaging over time scales
similar to $\tau_{\alpha}$ so as to filter out frequencies higher than
those that act over the periods of structural rearrangement.

We conclude this section by summarizing that we do not find a strong
$T$ dependence of the chains defined above. This is surprising because
if the highest forces (or any force chains) are related to slowing
down of dynamics in supercooled liquid, one might expect to see a
temperature dependence pattern in the chain properties.

\section{Fraction of localized and replaced particles in force chains}
\label{frac_loc_deloc}

\begin{figure}[tbp]
\begin{center}
\includegraphics[clip,width=15cm]{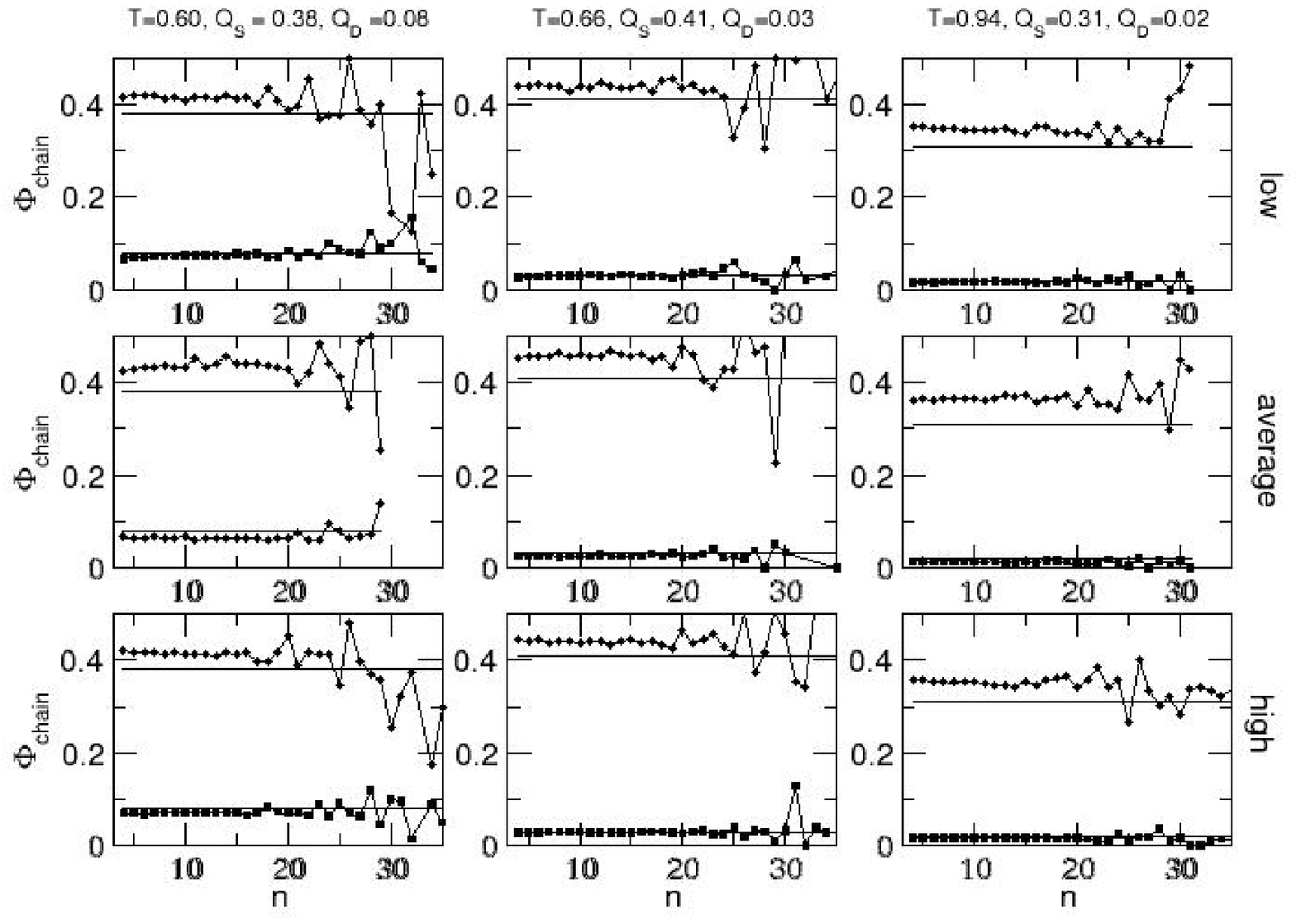}
\end{center}
\caption{Fraction, $\Phi_{{\rm chain}}$, of localized and replaced particles for low, average
and high force chains. The fraction of localized and replaced
particles at $t^{\rm{max}}_4$ is marked with the corresponding $T$, and
it is indicated by the solid lines on each panel.}
\label{chainsfrac}
\end{figure}

To look for a connection between force chains and mobility, we
calculate the fraction of localized and replaced particles in each
chain for all subsets of forces for which force chains are defined.
This fraction is calculated in the following manner. For each
configuration at $t$, we identify the force chains and localized and
replaced particles with respect to an initial configuration at $t=0$,
and for each chain of mass $n$ in the configuration at $t$ we count
the number of localized and replaced particles and divide by $n$.
After this, we average those fractions over all equally spaced
configurations. Figure~\ref{chainsfrac} shows the fraction of
localized and replaced particles in the chains of high, average and
low force at $t^{\rm{max}}_4$ (defined for each $T$ in
Ref~\cite{LacevicJCP2003}, and Section~\ref{summary_SHD_4pnt}) at
$T=0.60$, $T=0.66$ and $T=0.94$.We examine the configurations at the
particular time $t^{\rm{max}}_4$ because, as explained in
Section~\ref{summary_SHD_4pnt}, SHD as measured by $\chi_4(t)$ and
$\xi_4(t)$ is most pronounced then.  If SHD in supercooled liquids and
granular materials share common mechanisms, one might expect the
effects of jamming to be most detectable at this time. The fraction of
localized and replaced particles is shown in Figure~\ref{chainsfrac}
for each $T$.  The first column in Figure~\ref{chainsfrac} represents
$\Phi_{{\rm{chain}}}$ at $T=0.60$ for low, average and high force
chains.  The second and third column contain data for $\Phi_{{\rm
chain}}$ at $T=0.66$ and $T=0.94$ respectively.

From Figure~\ref{chainsfrac} we see that $\Phi_{{\rm chain}}$ is
essentially constant for $n \leq 20$ for both localized and replaced
particles.  For chain masses greater then $20$, $\Phi_{chain}$ becomes
noisy because of poor statistics. For chains with $n \leq 20$,
$\Phi_{{\rm chain}}$ has values slightly higher than the fraction of
localized particles in the bulk, $Q_S$ ($Q_S$ is indicated by solid
lines in Figure~\ref{chainsfrac}). This means that localized particles
are more likely to be found in chains with intermediate mass than
would be expected from the fraction of localized particles, ($Q_S$),
in the system.  $\Phi_{{\rm chain}}$ has values equal to or slightly
lower than the fraction of replaced particles in the bulk ($Q_D$ is
also indicated by solid lines in Figure~\ref{chainsfrac}). This means
that replaced particles are equally or less likely to be found in the
chains of intermediate mass based on their bulk population.  These
behaviors are basically universal regardless of the subset of forces
in which the force chains are defined.  Therefore, we cannot make a
simple connection with mobility, and force chains as defined here
probably do not play the same role as those found in granular
materials.

The population of localized and replaced particles in the force chains
can be explained if we suppose that localized particles reside in a
local environment that is less susceptible to force perturbations than
delocalized particles.  Imagine an "intrinsic" high force chain that
would be associated with the inherent structure and introduce a
perturbation near the chain that propagates perpendicular to it. This
perturbation can temporarily change which neighbors of a given
particle in the chain interact with it via the highest relative force,
thereby breaking the chain.  Such perturbations are the manifestation
of temperature in the system.  Since the localized particles are more
likely to be found in non-trivial chains, one might speculate that
something about their local environment makes them resistant to these
perturbations at least with respect to the relative forces exerted by
their neighbors.

\section{Discussion}
\label{discussions}

In this paper we examined the relationship between jamming in granular
materials and SHD in supercooled liquids.  We calculated the
instantaneous force distribution function $P(F)$ in a model
glass-forming liquid, and we did not find a peak in this quantity at
any $T$ whether above or below $T_g$, in contrast to the model
supercooled liquids studied in Ref.~\cite{O'Hern200186111}.  We also
found a possible connection between instantaneous force magnitude and
long term mobility. We defined force chains in our model glass-forming
liquid, and based on our results from Section~\ref{force_chains}, we
found force chains much longer than those that might be expected for
randomly assigned forces. These force chains probably do not play the
same role in this supercooled liquid as in granular materials because
they do not show a strong temperature dependence.

Force chains may, however, indicate a difference in the evolution of
the local environment of particles with different mobilities which may
be connected with cooperative and string like motion found in
model supercooled liquids close to $T_g$. Microscopic details of local
particles dynamics and the mechanism by which particles move along
string-like paths is studied in Ref.~\cite{YeshiJCP2004} in a
glass-forming Dzugotov liquid. The authors show that simultaneous
motion of individual particles along the string depends on the length
of the string, and that for shorter strings the motions is highly
coherent and for longer strings motion is coherent only within short
segments containing as many as seven particles (micro strings).  We
note the similarity between the anatomy of strings and force chains
studied here by comparing a snapshot of a string in Figure~10 of
Ref.~\cite{YeshiJCP2004} and Figure~\ref{chains26examp} of this paper
(remember that the criterion for a string is completely different from
the criterion for a force chain).  We also note that both the
distribution of mass of force chains and distribution of string length
appear to be exponential. Although the mass of the force chains does
not grow as a function of $T$, in contrast to string length, these
distributions are, to our knowledge, the only two distributions of
quantities associated with force and particle mobility measured in
supercooled liquids that obey exponential laws. This leads us to pose
certain questions: What is the relationship between force chains and
strings? Is it possible that force chain networks provide a
characteristic structural feature along which strings may occur?  Could
frequent defects in the force chain network due to breakage and
recombination of trivial force chains be associated with the local
liquid excitations proposed by Garrahan and Chandler~\cite{chandler},
and provide a path for string like motion?  Answers to those questions
may help us to understand the origin of SHD.  Further work on the
relationship between strings and force chains is ongoing.

\begin{acknowledgments}

We thank the National Partnership for Advanced Computing
Infrastructure (NPACI) program and the University of Michigan Center
for Advanced Computing for generous amounts of CPU time on the
University of Michigan AMD Athlon cluster. We thank Y. Gebremichael,
M. Vogel, J. W. Palko, M. W. Palko, C. S. O'Hern and M. O. Robbins for
useful discussions.

\end{acknowledgments}

\end{document}